\documentclass[prb,twocolumn,superscriptaddress,showpacs]{revtex4}

\usepackage{amsbsy}
\usepackage{amsfonts}
\usepackage{amssymb}
\usepackage{amsmath}
\usepackage{color}
\usepackage{graphicx}

\usepackage{dsfont}                  
\usepackage{braket}

\begin{document}


\title{Nonequilibrium localization and the interplay between disorder and interactions} 

\author{Eduardo Mascarenhas}
\affiliation{Institute of Theoretical Physics, Ecole Polytechnique F\'{e}d\'{e}rale de Lausanne EPFL, CH-1015 Lausanne, Switzerland} 

\author{Helena Bragan\c{c}a}
\affiliation{Departamento de F\'isica, Universidade Federal de Minas Gerais, C.P. 702, 30123-970, Belo Horizonte, MG, Brazil}

\author{R. Drumond}
\affiliation{Departamento de Matematica, Universidade Federal de Minas Gerais, C.P. 702, 30123-970, Belo Horizonte, MG, Brazil}

\author{M. C. O.  Aguiar}
\affiliation{Departamento de F\'isica, Universidade Federal de Minas Gerais, C.P. 702, 30123-970, Belo Horizonte, MG, Brazil}

\author{M. Fran\c{c}a Santos}
\affiliation{Departamento de F\'isica, Universidade Federal de Minas Gerais, C.P. 702, 30123-970, Belo Horizonte, MG, Brazil}

\begin{abstract}

We study the nonequilibrium interplay between disorder and interactions in a closed quantum system. We base our analysis on the notion of dynamical state-space localization, calculated via the Loschmidt echo. Although real-space and state-space localization are independent concepts in general, we show that both perspectives may be directly connected through a specific choice of initial states, namely, maximally localized states (ML-states). We show numerically that in the noninteracting case the average echo is found to be monotonically increasing with increasing disorder; these results are in agreement with an analytical evaluation in the single particle case in which the echo is found to be inversely proportional to the localization length. We also show that for interacting systems, the length scale under which equilibration may occur is upper bounded and such bound is smaller the greater the average echo of ML-states. When disorder and interactions, both being localization 
mechanisms, are simultaneously at play the echo features a non-monotonic behaviour indicating a non-trivial interplay of the two processes. This interplay induces delocalization of the dynamics which is accompanied by delocalization in real-space. This non-monotonic behaviour is also present in the effective integrability which we show by evaluating the gap statistics.


\end{abstract}


\maketitle
\section{Introduction}

The physics of transport in electronic systems is a cornerstone of our technologies and yet the general problem of non-equilibrium transport seems far from being completely understood. Significant steps in this direction have been taken starting with Anderson localization (or absence of diffusion)~\cite{Anderson}. It is well established that in the absence of interactions the energetic mismatch between neighboring sites in a lattice may completely prevent transport by localizing all single-particle states. In this scenario arbitrary weak disorder localizes one and two-dimensional systems, producing ideal insulators. Much less is known about the effects of the interplay between interaction and disorder.  
In fact, despite intensive studies over decades~\cite{DisInt1,DisInt2,Mirandinha,Vojta}, the  influence  of  electron-electron  interactions  on  the transport in disordered electronic systems is still a challenging problem. 

More recently it has been theoretically shown that the insulating phase survives the inclusion of a small amount of interaction in a disordered system although, in this case, one can induce transport by increasing the temperature ~\cite{FinteTMBL,FinteTMBL2}. This investigation has given rise to the emerging field of many-body localization (MBL), which is considered a dynamical transition; unlike an equilibrium (conventional) phase transition, in a MBL transition the properties of the entire set of eigenstates change and one cannot restrict the analysis to the ground-state physics. This dynamical localization has important consequences on time evolution, more specifically on the capacity of the system to self-equilibrate, that is, to act as a reservoir for 
its subsystems. Localized states may prevent equilibration, since local perturbations do not diffuse throughout the system - such systems are called non-ergodic~\cite{Gogolin}.

In the last few decades we have witnessed both experimental~\cite{MBLexp,Imma,ML} and theoretical~\cite{DMRG1,DMRG2,DMRG3} advances in the physics of one-dimensional quantum systems. In particular, after Basko's seminal work~\cite{FinteTMBL} many groups have focused on the non-equilibrium localization of interacting particles. In these works the MBL transition has been studied through different approaches, using, for instance, entanglement properties or eigenstate statistics. 
Numerical studies have shown that the spectral statistics of a one dimensional system crosses over from those of orthogonal random matrices in the diffusive regime at weak disorder to Poisson statistics in the localized regime at strong disorder~\cite{Huse1}. 
Concomitantly, dynamical features of disordered and interacting systems were observed focusing on multi-partite correlations~\cite{John} and on the entanglement entropy. 
It was shown that the interplay between interactions and disorder has a strong influence on the long-time behaviour of the entropy~\cite{BinaryD,EntGrowth,EntGrowth2,EntGrowth3,EntGrowth4}; later on it has been concluded that such signatures are just a characteristic trait of interacting systems, since these dynamical tendencies also exist in the absence of disorder and inhomogeneities~\cite{MBLsemDes}. An analysis of the spatial behaviour of the entropy (von Neumann entropy as a function of partition size) should be more appropriate to address localization, since it saturates when the partition size reaches the localization length~\cite{Berko}. However this is a difficult approach, since the investigation has to be applied to large chains. In fact, the definite evidence of a MBL transition remains elusive despite creative and innovative efforts~\cite{Huse2,Rossini,ErgoEcho,PollmannDiag,Huse,AnoGriff,DEER,Luitz,Hauke,BinaryD} due to technical difficulties, such as finite size limitation of numerical 
calculations.

In this work we study the dynamical state-space localization of an initial state that is maximally localized in real-space and evolves in time with an interacting and disordered Hamiltonian.
In this way we are able to describe the non-equilibrium localization effects of disorder and interaction in a well defined way which is accessible for small chains. 
Furthermore, we propose a direct relation between dynamical localization in the state-space and the more standard real-space localization perspective.  
We show that when acting alone both disorder and interaction progressively and monotonically localize the system. However, when combined, they may lead to delocalization.

The paper is organized as follows. Section~\ref{II} is dedicated to the model. We study a one-dimensional (1-D) spinless fermion system with nearest-neighbor Coulomb interaction and on-site disorder. The quantity we choose to address dynamical state-space localization (the Loschmidt echo) is described in section~\ref{III}, as well as the connection with real-space localization. We present our numerical results in section~\ref{IV}. The last section is dedicated to our comments and conclusions. 

\section{Model}\label{II}

The disordered Anderson-Hubbard model is a standard Hamiltonian used to describe the competition between kinetic energy, electron-electron interaction, and disorder.
The fully polarized Anderson-Hubbard model, or a disordered spinless fermion model, can still describe this competition with less computational effort. In the latter, we only retain the nearest-neighbor Coulomb interaction since a site cannot be doubly occupied. The Hamiltonian is then given by
\begin{eqnarray}H=\sum_{\langle i,j\rangle} \left[ -\frac{J}{2}(c^{\dagger}_ic_{j} -c_ic^{\dagger}_{j})  +\epsilon_in_{i} \right. \nonumber \\ \left. +U(n_{i}-1/2)(n_{j}-1/2)\right].\label{eqHami}\end{eqnarray}
The first term describes the kinetic energy, with $J$ being the tunneling between nearest neighbors; the second describes the local energy of the $i$th site, with $\epsilon_i$ taken from
a uniform distribution $[-W,W]$; finally, the last term represents the nearest-neighbor Coulomb repulsion, with $U$ being
the interaction strength.
 $c_i$ and $n_i$ are the destruction and number operators acting on site $i$, respectively. Throughout this paper we use $J=1$ as the unit of energy. 
This model can be mapped to the XXZ spin-1/2 chain through a Jordan-Wigner transformation~\cite{SpinToFermion}. 

In this work we study small chains with $N=10,12,14,16$ sites and solve the Hamiltonian by exact diagonalization. Although the study of small chains is not ideal to describe phase transitions (defined in the thermodynamic limit), it can still produce relevant physical insights about the interplay between interactions and disorder. Furthermore, small systems can be realized in chains of trapped ions~\cite{ions} and cold atoms~\cite{Imma} and have been the subject of many recent theoretical studies~\cite{EntGrowth,EntGrowth2,EntGrowth4,Huse2,John}. 

\section{Nonequilibrium Localization}\label{III}

We choose to describe the competition between disorder and interactions through a state-space localization perspective. 
The \emph{dynamical state-space localization} can be described by the inverse participation ratio (IPR)~\cite{ErgoEcho} of the time average ensemble.
The IPR is calculated through the purity of the time-average ensemble, which gives us an estimate of the inverse of the effective length or area covered by the dynamics:
\begin{equation}IPR=\mathrm{tr}\{ \bar{\rho}^2 \},\quad \bar{\rho}=\lim_{T\rightarrow\infty}\frac{1}{T}\int_0^T |\Psi_t\rangle\langle\Psi_t|dt,\end{equation} 
where $|\Psi_t\rangle$ represents the dynamically evolving state at a time $t$, starting from an initial state $|\Psi_0\rangle$.
The IPR is bounded between unity and $1/d^N$ with $d$ being the site dimension and $d^N$ the dimension of the state-space. If $IPR=1$ the dynamics are maximally localized meaning that an initial pure state is unaltered throughout the evolution; if $IPR=1/d^N$ the dynamics are maximally delocalized in the sense that the time-average ensemble is the identity ensemble in which all states contribute with the same probability. 

The purity of the average ensemble can also be expressed by the time-average of the Loschmidt echo $L(t)$
\begin{equation}IPR=\bar{L}=\lim_{T\rightarrow\infty} \frac{1}{T}  \int_0^T dt\  L(t)\label{time}\end{equation} where $L(t)=|\langle \Psi_0| e^{-iHt}|\Psi_0\rangle |^2$.
This relation becomes useful in the case where the full diagonalisation or long-time dynamics are unreachable (which is the case experimentally~\cite{MBLexp}) and we may use the dynamical behaviour of the echo as a figure of merit. 
Furthermore, the echo has proven to be an important tool in characterizing several aspects of the closed system dynamics~\cite{ELea,ELea2,ELea3,ELea4}. 

Considering a non-degenerate Hamiltonian, the above relation corresponds to  
\begin{equation}\bar{L}= \sum_E|\langle E|\Psi_0\rangle |^4,\label{echo}\end{equation} 
where $|E\rangle$ represents the Hamiltonian eigenstates. In this way, the time averaged echo is related to the fidelity between the initial state and the Hamiltonian eigenstates. 

In this work we calculate the dynamic state-space localization through eq. (\ref{echo}). In summary, the dynamics of an initial state allowed to evolve in time with a given Hamiltonian is localized in the state-space when the participation ratio is close to unity ($1/\bar{L} \approx 1$), while it is delocalized if $1/\bar{L} \approx d^N$, meaning that the state covers a large portion of the state-space through the time evolution. 
\subsection{Connection between state-space and real-space localization}

The \emph{equilibrium real-space localization}, as introduced by Anderson~\cite{Anderson}, is a single-particle effect in which the wavefunction exhibits exponentially decaying tails. In this way, each particle can only be found, with considerable probability, inside a finite localization length $\xi$ from a given lattice site. The definition of real-space localization in a many-body system is more subtle~\cite{Gogolin} and in general we would refer to a correlation length rather than a localization length. 
Here we relate state-space localization with real-space localization by investigating the participation ratio of the dynamics of a special set of initial states, namely half-filled random \emph{product} states, i.e. states of the type $|\Psi_0\rangle=\prod_{i=1}^N |n_i\rangle$ and $\sum_{i} n_i=N/2$, where $|n_i\rangle=\{|0\rangle, |1\rangle\}$ are the eigenstates of the number operator of the ``i-th'' site, and $N$ is the number of sites in the chain. Note that such states represent electrons which are \emph{maximally localized} in real-space, with a vanishing correlation length ($\xi\rightarrow 0$), and we will name their set $\{|E_{\xi\rightarrow0}\rangle\}$.
We perform an average of  $\bar{L}$ over $M$ initial product states and disordered Hamiltonian distributions
\begin{equation}
 \langle\langle \bar{L} \rangle\rangle=  \frac{1}{M}\int \prod_id\epsilon_i p(\epsilon_i)\sum_{E^{'}}^M \sum_E |\langle E(\{\epsilon\}) | E^{'}_{\xi\rightarrow0}\rangle |^4\le 1, \label{average}
\end{equation}
where $\langle\langle \ \rangle\rangle$ denotes the total averaging, $\{\epsilon\}$ represents the set of drawn on-site energies and $\prod_i p(\epsilon_i)=P(\{\epsilon\})$ is the probability of a disorder configuration.

In the extreme case in which a given product state $|E^{'}_{\xi\rightarrow0}\rangle$ is allowed to evolve in time
with a Hamiltonian whose eigenstates $\{|E_{\xi\approx0}\rangle\}$ are strongly localized in real-space (i.e. very weakly correlated), it 
will necessarily have a high projection on one of them, i.e. $|\langle E_{\xi\approx0} |E^{'}_{\xi\rightarrow0}\rangle |^2 \approx \delta_{E,E^{'}}$.
In this way, the state almost does not change in time, covering a
small portion of the Hilbert space during the time evolution
(state-space localization), resulting in $1/ \langle \langle \bar{L} \rangle \rangle \approx 1$.
On the other hand, if the eigenstates of $H$ are $\{|E_{\xi\gg1}\rangle\}$ delocalized in real-space,
$|E^{'}_{\xi\rightarrow0}\rangle$ will have a small projection all of them, $|\langle E_{\xi\gg1}|E^{'}_{\xi\rightarrow0}\rangle |^2\ll1$. Therefore, the state covers
a large portion of the Hilbert space during the dynamics,
resulting in a large participation ratio $1/ \langle \langle \bar{L} \rangle \rangle \gg 1$ (dynamical state-space delocalization). 
In general, we expect that, when using random
product states (which are \emph{maximally localized} with vanishing correlation length) as initial states the dynamical localization
becomes closely connected to the real-space localization of the Hamiltonian eigenstates. Therefore, we expect a functional dependence $\langle\langle \bar{L} \rangle\rangle\propto f(\xi) $, such that $f(\xi)$ is a decaying function of $\xi$. 

It is important to emphasize that the direct relation between real-space and state-space localization depends on the choice of initial state. Indeed, there are cases where the states are delocalized in real-space but localized in state-space. The simplest example is to consider initial states that are (at least approximately) eigenstates of a delocalized Hamiltonian. In this case the state would have an extremely long correlation length, however, the echo would remain (almost) unaltered throughout the evolution.


\subsection{Average participation ratio of maximally localized states and single particle localization}
One example where $f(\xi)$ can be explicitly obtained is that of noninteracting particles in disordered systems at low density. In this regime we may approximate the problem by the single particle analysis which allows for a direct analytical treatment of the quantum states. 
The presence of scattering centers in a potential induce bound eigenstates which are exponentially decaying around localization centers at $X$ as $|E_{\xi}(X)\rangle=\int dx \frac{1}{\sqrt{\xi}}e^{-|x-X|/\xi}|x\rangle$, with a localization length $\xi$ that is proportional to the depth of the potential. 

In our analysis we always initiate the system in a maximally localized state which in the single particle case translates as $|x'\rangle=\lim_{\xi\rightarrow0}|E_{\xi}(x')\rangle$. Let us now evaluate the echo dependence on the localization length for this particular case:
\begin{equation} \langle\langle\overline{L}\rangle\rangle\propto \frac{1}{V}\int dx'\int dX\left|  \langle E_{\xi}(X)|x'\rangle \right|^4=\frac{1}{2\xi},\end{equation}
showing that the average echo is inversely proportional to the localization length and $f(\xi)=\frac{1}{\xi}$ as we expected from the reasoning of the last section. Therefore, if the Hamiltonian has very deep effects inducing a vanishing localization length the dynamics is localized in state-space. In the other limit when the Hamiltonian is almost clean with very large localization length the dynamics is highly delocalized. Hence, we establish an explicit relation between the state-space dynamical localization and the real-space localization. Certainly, such simple functional relation is not necessarily expected for interacting particles.

\subsection{Average participation ratio and equilibration}

The previous reasoning can also be seen from the equilibration perspective that allows for the analysis of a broader class of Hamiltonians which may account for interactions. The fact that the dynamics of the initial states (which are chosen to be localized in real-space) are localized in the Hilbert space implies in the breakdown of the ergodic hypothesis; the value of a time averaged observable depends on the initial state and the system does not equilibrate. The absence of equilibration has been used as an indicator of real-space many-body localization in both experimental~\cite{MBLexp} and theoretical~\cite{small1} studies. 
%

More specifically, if one splits a system of $N$ sites in two parts ($S$ and $R$) and chooses two different initial product states $|\pi_S\rangle\otimes|\pi_R\rangle$ and $|\tilde{\pi}_S\rangle\otimes|\pi_R\rangle$ (with $|\pi_S\rangle$ orthogonal to $|\tilde{\pi}_S\rangle$), equilibration guarantees that this choice becomes undetectable, for any practical purpose, if given enough time. This means, in particular, that the time average of the trace distance $D(\rho,\tilde{\rho})=\frac{1}{2}||\rho_{S}-\tilde{\rho}_{S}||_{1}$ of the reduced evolved states of $S$, $\rho_S(t)$ and $\tilde{\rho}_S(t)$, should approach zero as $t\rightarrow \infty$: $\lim_{T\rightarrow\infty}\frac{1}{T}\int_{0}^{T}D(\rho_{S}(t),\tilde{\rho}_{S}(t))dt \rightarrow 0$. On the other hand, the distance remains large in the absence of equilibration. In the appendix, we prove that if $1/\braket{\braket{\overline{L}}}<1+\delta$, with $\delta \ll1$ related to the size $N_S$ of the subsystem $S$ as $d^{-N_S} \leq \frac{\delta}{1+\delta}$, 
then $$\lim_{T\rightarrow\infty}\frac{1}{T}\int_{0}^{T}D(\rho_{S}(t),\tilde{\rho}_{S}(t))dt\geq 1-6\sqrt{\delta}.$$ 

It is thus shown that if the dynamics of the global system is localized in state-space then it is guaranteed that large enough subsystems composed of $N_S$ sites do not
equilibrate. This means that correlations do not extend
over $N_S$ sites and thus the correlation length should be
much smaller than this subsystem.

\section{Numerical Results}\label{IV}
\begin{figure}
\begin{center}
\end{center}
\begin{center}
\includegraphics[width=07cm]{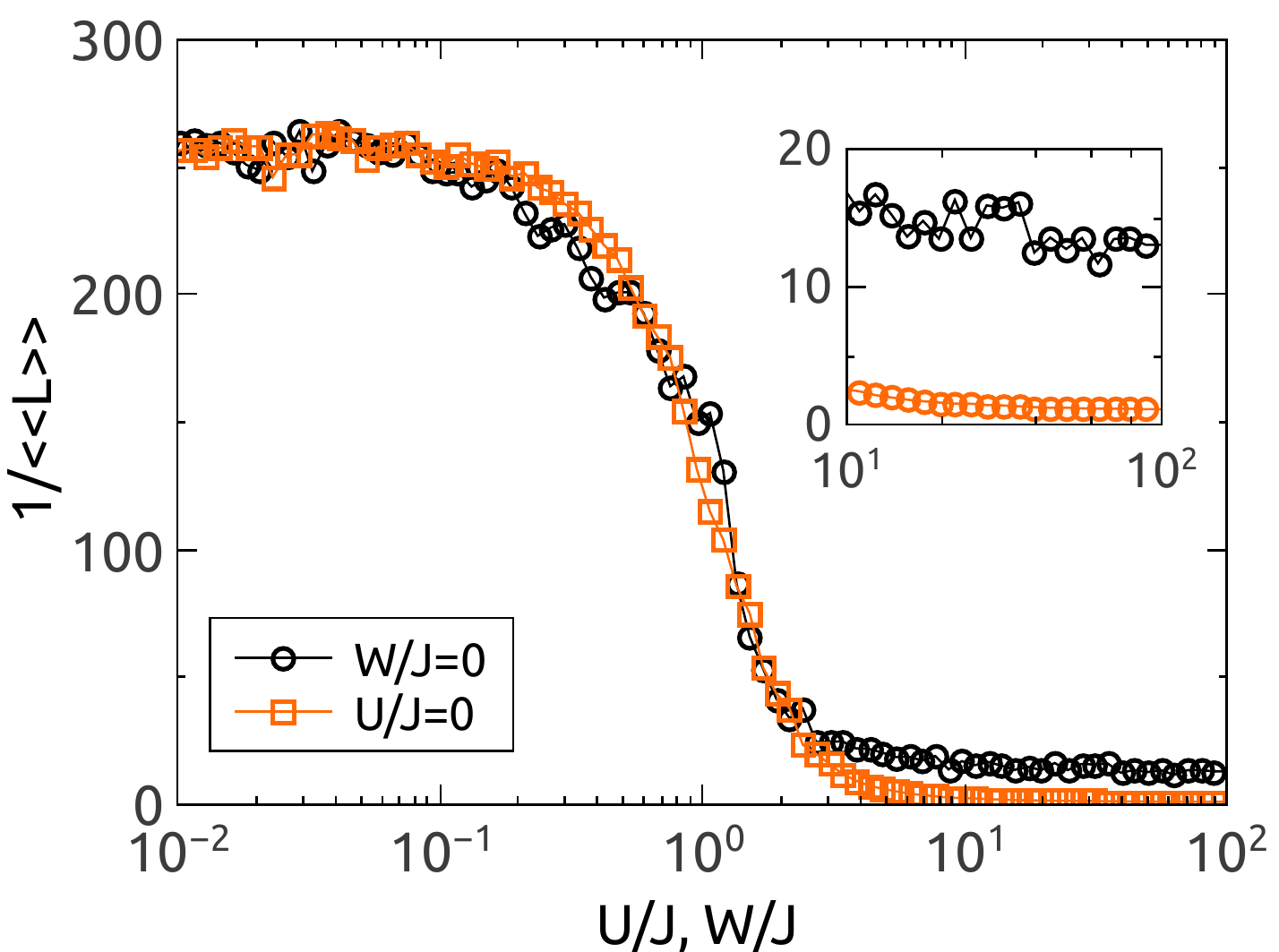}
\caption{(Color online) Participation ratio ($1/ \langle \langle  \bar{L} \rangle \rangle$) of maximally localized states evolving with a non-disordered Hamiltonian as a function of interaction $U$ (black circles) and with a non-interacting Hamiltonian as a function of disorder strength $W$ (red squares). The inset shows a zoom for large interaction/disorder. Results obtained for a system of $N=12$ sites through an average over $10^4$ initial states and disorder realizations.} 
\label{Fig1}
\end{center}
\end{figure}

We start by evaluating the participation ratio, $1/\langle\langle \bar{L} \rangle\rangle$, through eq. (\ref{average}) for the two simplest cases, that is, using a disordered Hamiltonian in the absence of interaction and also an interacting non-disordered Hamiltonian. We perform an average oven $10^4$ maximally localized initial states and, in the disordered case, for each initial state we randomly select each
site energy with uniform probability between $-W$ and $W$. In the absence of interactions the presence of disorder localizes the system progressively decreasing its localization length. This result is represented in Fig.\ref{Fig1} by the monotonic decrease of the participation ratio as we increase disorder, which means the purity of the average ensemble only increases. In this case the decrease of the participation ratio follows the reduction of transport due to Anderson localization.

Similarly, for a non-disordered Hamiltonian, interaction progressively localizes the system. However the strongly interacting system is less localized than the strongly disordered system. The state-space localization analysis seems to indicate a partial transport suppression due to interaction. This is in agreement with the equilibrium counterpart with a non-null conductivity in a system with large interaction but finite temperature~\cite{ResCond}.

The difference between both interaction and disorder routes to localization shown in the inset of Fig. \ref{Fig1} can be understood by applying standard time-independent perturbation theory to the model Hamiltonian of equation (\ref{eqHami}). For a specific disorder realization, the disordered term of the Hamiltonian is diagonal in a basis of random product states and the spectrum of a finite system is non-degenerated.
Thus, for strong disorder any product state tends to an eigenstate of the system. In the limit of non-interacting and strongly disordered one-dimensional systems \textit{all} the eigenstates of the Hamiltonian (\ref{eqHami}) are strongly localized; this result leads to $1/ \bar{L} \approx 1$.  

The electron-electron interaction term of the Hamiltonian (\ref{eqHami}) is diagonal in the same basis. Its eigenstates, however, are degenerated in sub-spaces of equal number $m$ of nearest-neighbor filled sites, which we will refer to as $U_m$ subspaces. By including the hopping via degenerated perturbation theory we break the energy degeneracy of these eigenstates and the new eigenstates of the system become superpositions of localized states belonging to the same $U_m$ subspace. The dynamics of a given initial product state with $m$ nearest-neighbor filled sites is restricted to the corresponding $U_m$ subspace being, therefore, partially localized. That is why, in the strongly interacting limit, we still observe some residual dynamics.


\begin{figure}
\begin{center}
\end{center}
 \begin{center}
\includegraphics[width=07cm]{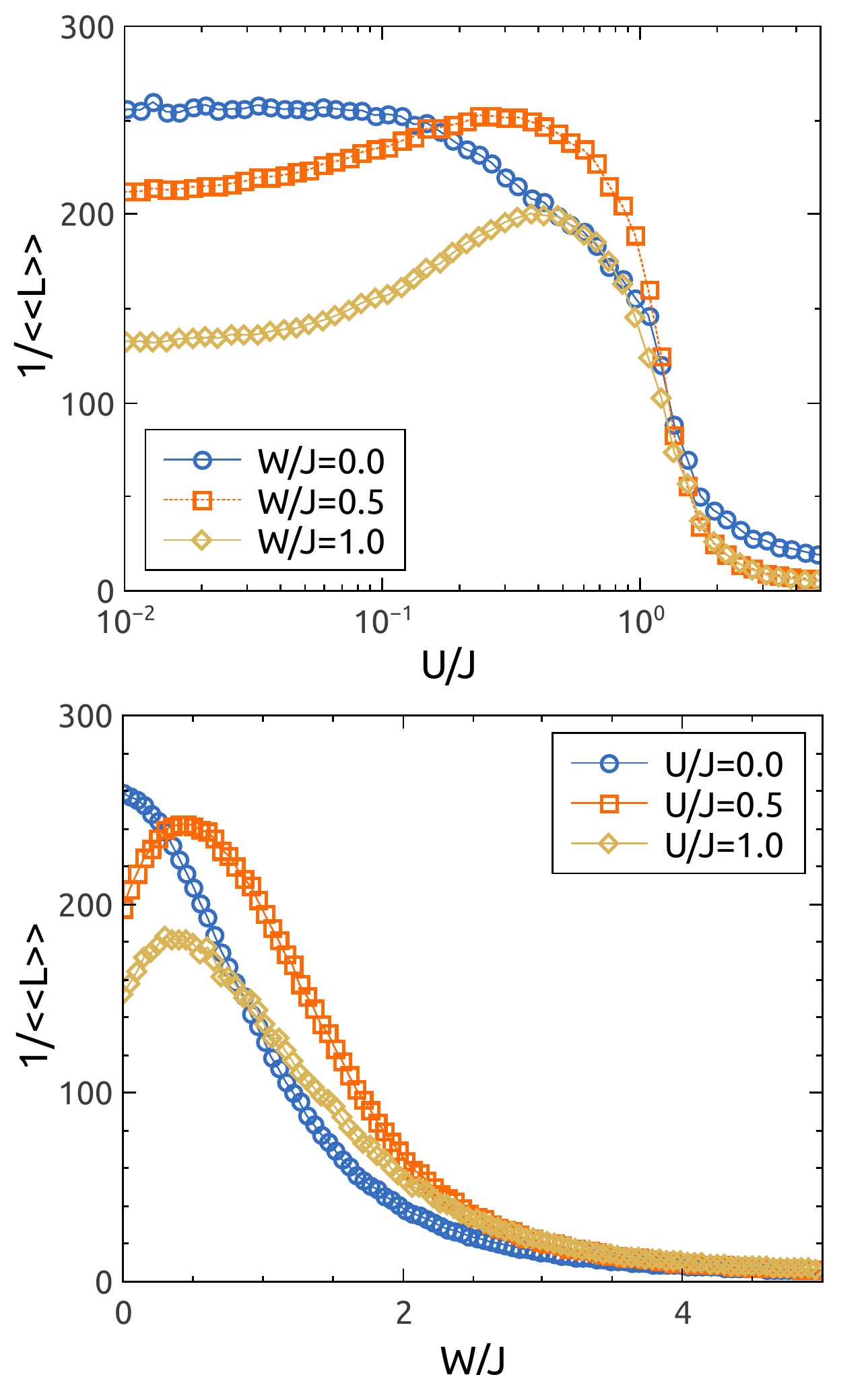}
\caption{(Color online) Participation ratio, $1/\langle \langle \bar{L} \rangle \rangle$, for (top) disorder strengths $W$  as a function of interaction $U$, and (bottom) different interactions $U$ as a function of disorder $W$. The results were obtained averaging over $10^4$ different initial random product states and disorder realizations for a chain of $N=12$ sites.} 
\label{Fig2}
\end{center}
\end{figure}

The competition between disorder and interactions leads to a challenging scenario which can not be described through perturbation theory, but the effects can still be observed in the participation ratio, as can be seen in Fig. \ref{Fig2}. Even though their individual natural effect is to localize the system, the combination of disorder and interactions leads to nontrivial behaviour. In fact, the introduction of disorder in an interacting system (top panel) or of interactions in a disordered system (bottom panel) leads to delocalization. Such interplay continues up until the point at which one of the effects starts to dominate, i.e. when either disorder or interaction becomes the dominant effect. More specifically, the interplay between interaction and disorder produces a tendency to delocalization when both $U$ and $W$ are of the same order as the hopping 
amplitude $J$. These results are imprinted in the non-monotonic behaviour of the participation ratio. In the thermodynamic limit this behaviour could lead to a reentrant phase diagram similarly to the infinite temperature equilibrium diagram outlined in~\onlinecite{InfReentrant} and the non-disordered non-equilibrium diagram outlined in~\onlinecite{NonDisReent}, noting that in the present case it is a genuine non-equilibrium effect of the interplay between interactions and disorder.

\begin{figure}
{\includegraphics[width=7cm]{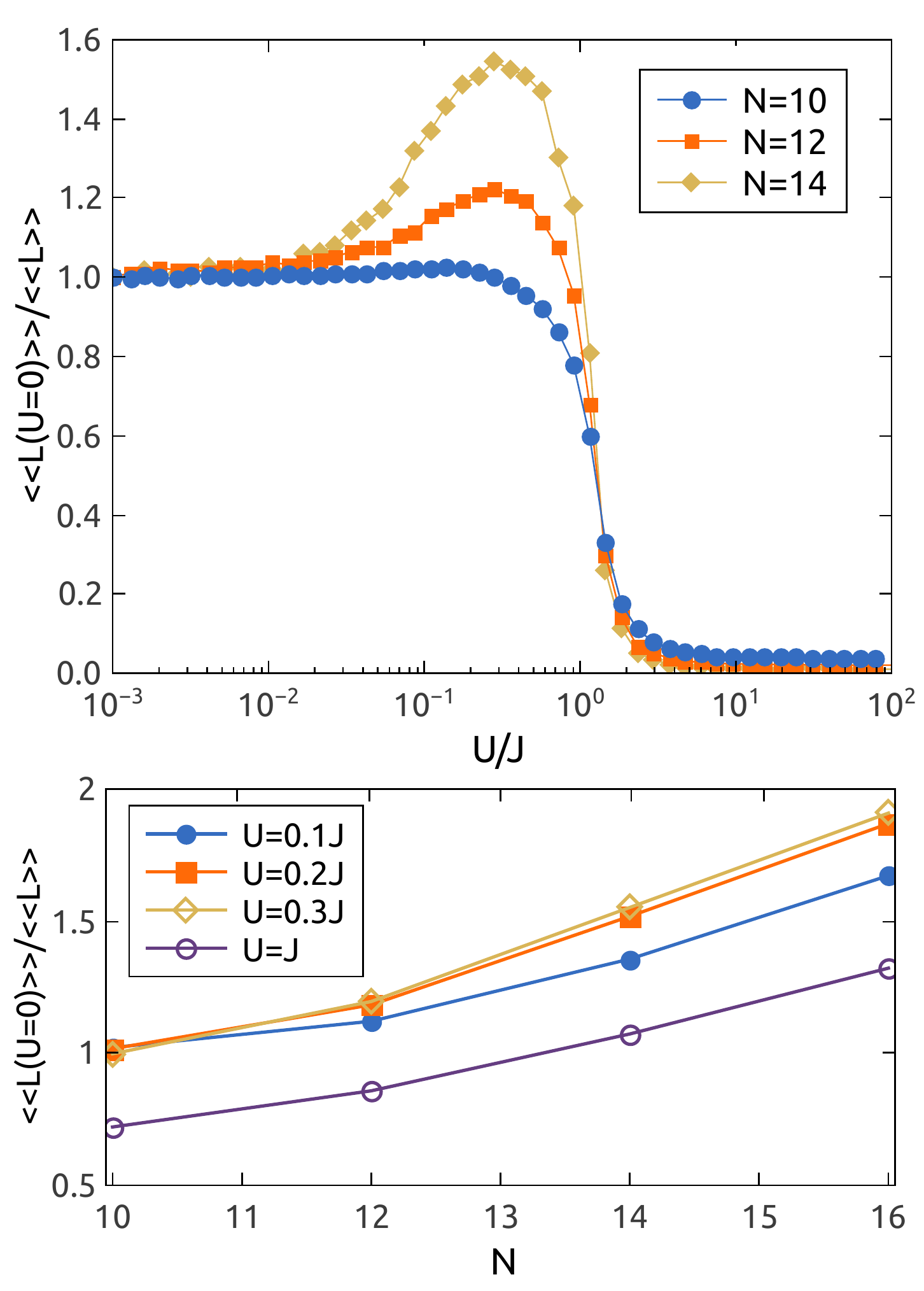}
\caption{Scaled inverse echo, $\left(\langle \langle \bar{L}(U) \rangle \rangle/\langle \langle \bar{L}(U=0) \rangle \rangle \right )^{-1}$ for $W=0.5$ (Top) as a function of interaction strength and (Bottom) as a function of the number of sites. Results averaged over 5000 samples for $N=10$, 2000 samples for $N=12$, 1000 samples for $N=14$, and 500 samples for $N=16$.}  
\label{ManyN}}
\end{figure}

The nonequilibrium interplay between disorder and interactions
becomes more pronounced for larger systems. This is evident in Fig.~\ref{ManyN} where we show the rescaled inverse echo for different system sizes as a function of the interaction strength. We scale the echo by the value it assumes when the interaction vanishes. In that case when $\langle\langle \bar{L}(U=0)\rangle\rangle/\langle\langle \bar{L}\rangle\rangle>1$ the interplay induces delocalization. In the top panel of Fig.~\ref{ManyN} we see that the region and magnitude above unity increase with the system size. From the bottom panel we confirm that this effect becomes stronger for larger systems. We point out that the total mutual entropy has also been shown to have a non-monotonic behaviour~\cite{John}. This, once again, hints on the fact that the interplay is also manifested in the spacial correlations and that, indeed, defined with respect to maximally localized states, the average echo is a useful quantifier of localization out of equilibrium.
We note that all the curves for different $N$ cross at the same point and we could argue that such crossing could indicate a phase transition in the same spirit that has been previously done in the literature for other quantities. However, we feel that such a claim is unsubstantiated in the sense that such chain sizes are too small for extracting information about the correlation length in the thermodynamic limit.

\begin{figure}
{\includegraphics[width=7.5 cm]{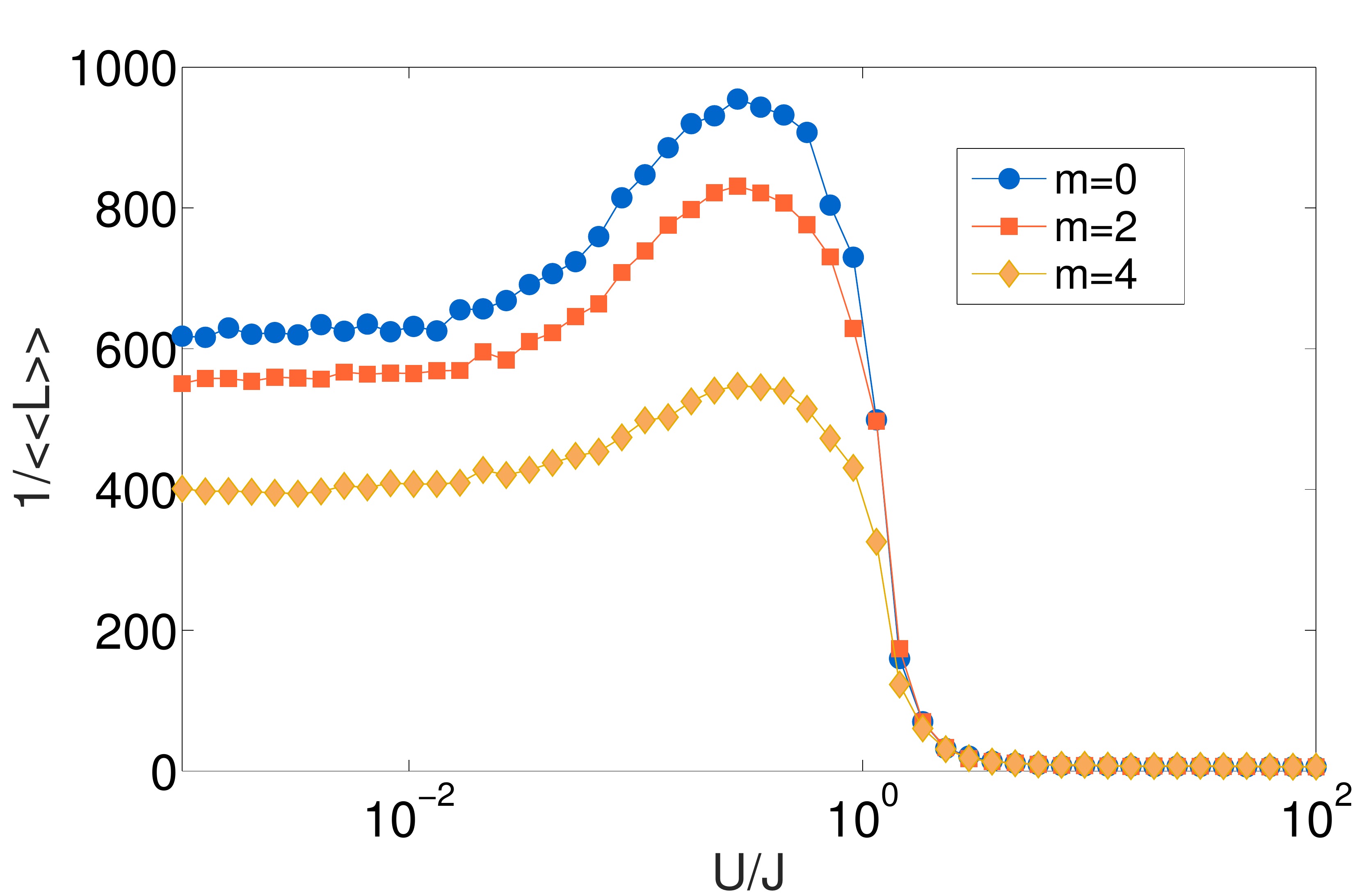}
\caption{Inverse echo for $W=0.5$ as a function of interaction strength for different magnetization subspaces. Results averaged over 1000 samples for $N=14$.}  
\label{mag}}
\end{figure}

Figures~\ref{Fig1} to~\ref{ManyN} exhibit results for a subspace of half-filled chains (zero magnetization subspace, in the spin scenario); in Fig.~\ref{mag} we show the inverse echo for different filling sectors, analyzing how the results depend on the filling fraction, that is, the number of particles over the number of sites. Add one and two particles to a system at half-filling (corresponding to subspaces of magnetization $m=2$ and $m=4$, in spin units, respectively) progressively localizes the system, as indicated by the reduction of the participation ratio. The non-trivial interplay between interaction and disorder, which leads to a non-monotonic behavior of the inverse echo, is still present in the sectors we have analyzed, however it becomes less pronounced. In fact, in the limiting case of fully occupied or fully empty system (maximal magnetization) disorder and interactions would not play any role in the participation ratio.

Our results should be distinguished from the results in~\onlinecite{InterplayDemintirinha}, in which a local echo is defined based on the magnetization. Firstly the non-monotonic behaviour in~\onlinecite{InterplayDemintirinha} is already present both at zero disorder (and varying interaction) or zero interaction (and varying disorder); secondly even though local probes are highly desirable experimentally, this one in particular is not a direct measure of localization, although a reasonable indicator in some cases. 
We also point out that our findings for the Loschmidt echo differ significantly from those in~\onlinecite{ELea5} for several reasons: there, the authors use a class of initial states that are local representatives of an energy shell, while our choice is not based on energy but rather randomly selected from the entire set of non-entangled states. Their focus is also on the Heisenberg point while we sweep the interaction strength.

\begin{figure}
\begin{center}
\end{center}
 \begin{center}
\includegraphics[width=7cm]{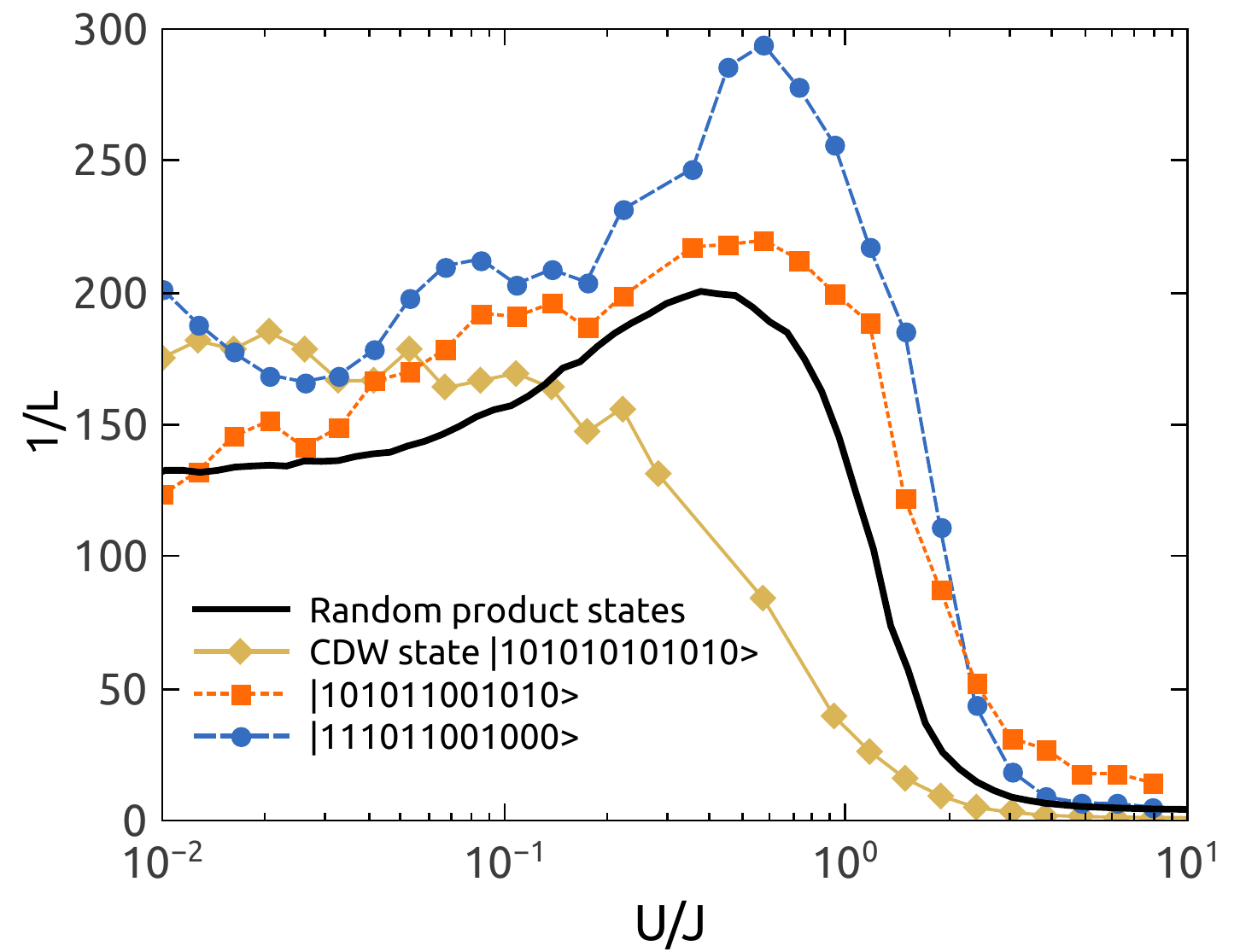}
\caption{(Color online) Participation ratio for $W=1$ obtained starting from specific half-filled initial product states. The solid black line corresponds to the data of Fig. \ref{Fig2}, i.e., $\langle \langle \bar{L} \rangle \rangle$ is obtained through an average over $10^4$ initial product states and simultaneous disorder realizations. In the other curves, we start from the initial states specified in the legend and average $\bar{L}$ over $10^3$ disorder realizations. Results for $N=12$.} 
\label{Fig3}
\end{center}
\end{figure}

As described above, we perform an average over different initial random product states of full or empty sites, resulting in a half-filled chain (we remember that the model can be mapped onto a spin-half XXZ chain, and in the spin scenario this set of states is equivalent to random product states of spins up and down, with null magnetization). This set is a good representative of real-space localized states in general, which are strongly non-entangled.
Our choice of initial states is essential to the observation of our results and to the assurance of our interpretations, as confirmed by the results of Fig. \ref{Fig3}.
Performing the same analysis described above, but starting with a specific local state, one could obtain different results. Using a  charge density wave state (CDW state, $|101010...\rangle$) as the initial state, for instance, we observe a monotonic decrease of the participation ratio as we increase interaction in a disordered system (see Fig. \ref{Fig3}). It happens because the CDW state, which is the ground state of the strongly interacting clean system, is not a typical localized state. It is contained in the $U_0$ (no nearest-neighbor-filled sites) subspace, spanned by only 2 of the 924 possible product states in a half-filled chain. Fig. \ref{Fig3} also shows that 
starting with a state that is a simple site-permutation of the CDW state is enough to recover the non-monotonic behaviour.     

\begin{figure}
\begin{center}
\end{center}
 \begin{center}
\includegraphics[width=07cm]{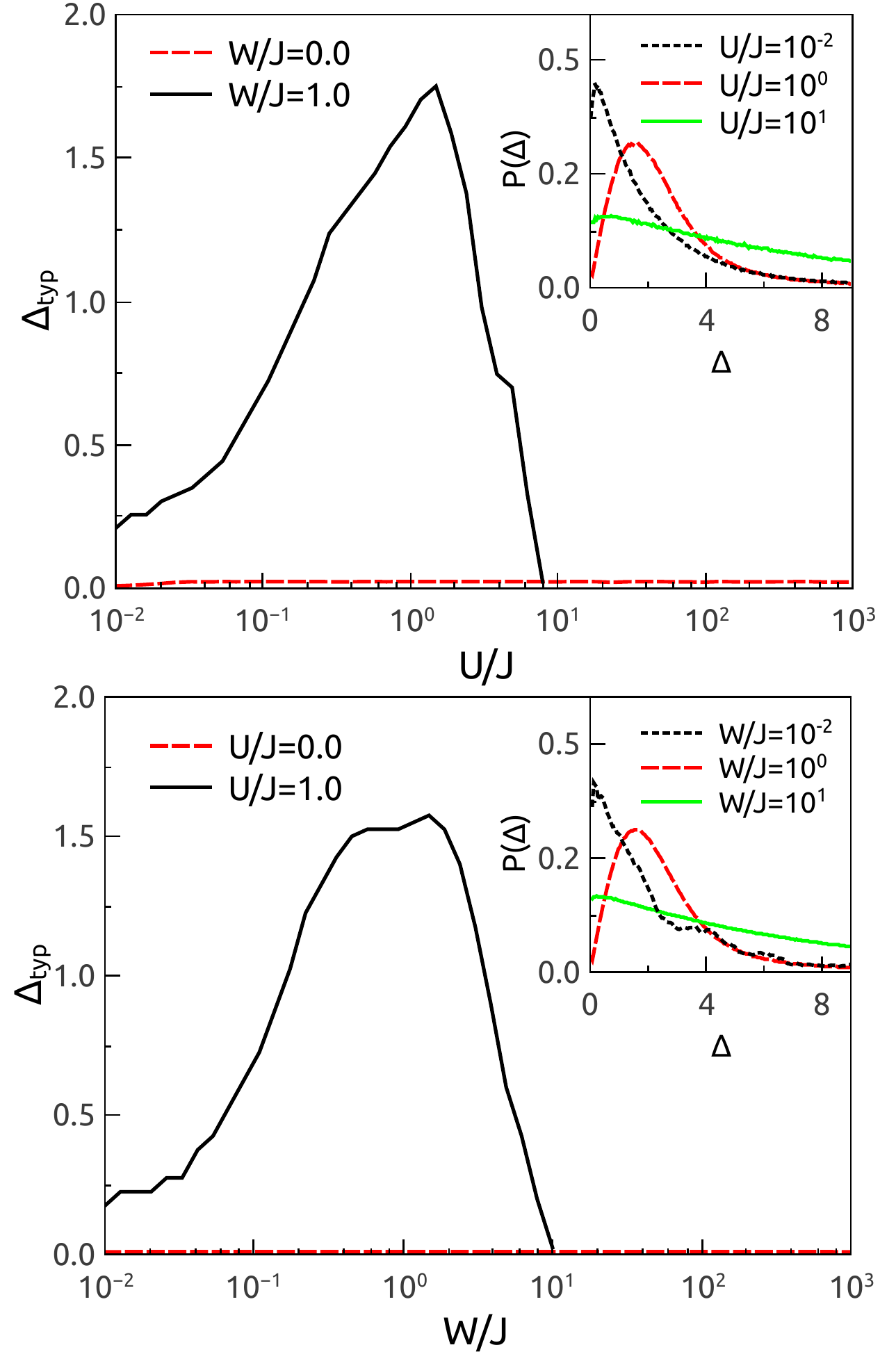}
\caption{(Color online) Typical gap, given by the position of the maximum of the gap distribution, as a function of $U$ (top) and $W$ (bottom). The inset in the top panel shows the gap distributions for a system with $W=1$ and different values of $U$, while the bottom inset shows the distributions for $U=1$ and different disorder strengths. Results for $N=12$ and averaged over $5 000$ (top) and $20 000$ (bottom) disorder realizations.} 
\label{Fig4}
\end{center}
\end{figure}

To complete the analysis, we show that the interplay between disorder and interactions is also manifested in the gap distribution of the Hamiltonian, $P(\Delta)$, where $\Delta=E_{i+1}-E_i$ and $\{E_i\}$ are the Hamiltonian eigenvalues in ascending order. 
The gap distribution is commonly used as an indicator of integrability or chaos. Generically, integrable system exhibits Poissonian gap statistics while chaotic systems exhibit Wigner-Dyson statistics~\cite{ChaosBooks}. 
Furthermore, we expect localized regimes to resemble integrable system and their statistics (since they have local constants of motion) while delocalized regimes are expected to resemble non-integrable systems~\cite{Cross}. A substantial difference between these two regimes is that the maximum of the distribution moves closer to zero in case of localization and to positive values in case of delocalization. In Fig.~\ref{Fig4} we keep track of the typical gap, $\Delta_{typ}$, given by the position of the maximum of the distribution (examples of $P(\Delta)$ for disordered and interacting systems are given in the insets). This figure clearly shows that the gap statistics of a clean interacting system and of a disordered non-interacting system is always a Poisson-like distribution since they are integrable. On the other hand, the inclusion of 
disorder in an interacting system or the inclusion of interaction in a disordered system, leads to a non-monotonic behaviour of the typical gap, indicating  that the nontrivial interplay between disorder and interactions drives the system to a more delocalized regime when $W\approx U \approx J$. $\Delta_{typ}$ moves closer to zero as we increase disorder or interaction even further, recovering the localized regime. This analysis agrees with our previous results, that is, the maximum of $\Delta_{typ}$ as a function of disorder or interaction is established for the same range of parameters in which we observe the maximum of the participation ratio (Fig. \ref{Fig2}).

\section{Conclusion}\label{VI}
We have used the time-averaged Loschmidt echo of maximally localized states to characterize state-space localization. We applied this concept to show that for a single particle the average echo is inversely proportional to the localization length, highlighting an intimate relationship between dynamical state-space localization and real-space localization. We have then extended the idea to many-body localization where we have derived a bound for subsystem equilibration based on this echo. 
Using this approach, we have shown that in a system featuring disorder and repulsive interaction, both localizing mechanisms, the interplay between them can actually lead to delocalization. This is a direct consequence of the fact that each mechanism localizes the dynamics in a different sense and, when they are of the same order of the hopping amplitude, their competition partially cancels each other's effect. This non-trivial interplay is also imprinted in the gap distribution of the Hamiltonian indicating that it also affects the integrability of the system. Finally, the interplay is shown to become more pronounced with increasing system size, robustly showing at least partial disorder-interaction cancellation.

We thank Rodrigo G. Pereira for useful discussions and Gabriele de Chiara for providing a code with which part of the results were obtained. We also acknowledge Gisele I. Luiz for obtaining some of the numerical data presented in this paper. This work was supported by CNPq, CAPES and FAPEMIG. Computations were partially performed in the CENAPAD-SP.

\onecolumngrid
\appendix

\section{Equilibration analysis}

\textbf{Proposition}: Let $\mathcal{H}=(\mathds{C}^{d})^{\otimes N}$ be the state space of a $N$ sites quantum system
with non-degenerate Hamiltonian $H$. Assume
$1/\braket{\braket{\overline{L}}}<1+\delta$
, with $\braket{\braket{\text{ } }}=\frac{1}{^{d^{N}}}\sum_{\pi}$ where the sum runs over the elements $\ket{\pi}$ 
of some fixed product basis $\Pi$,
$\overline{L}(\ket{\pi})=\sum_{E}|\braket{\pi|E}|^{4}$, with the sum running over all 
eigenstates $\ket{E}$ of $H$, and
$0<\delta<1$. For every subsystem $S$ with $N_{S}$ sites satisfying
$d^{-N_{S}}\leq\frac{\delta}{1+\delta}$, there exists a pair of 
orthogonal product states
$\ket{\pi_{S}}$ and $\ket{\tilde{\pi}_{S}}$ on $S$ and a product state 
$\ket{\pi_{R}}$ on the remaining system
$R$, such that
$$\lim_{T\rightarrow\infty}\frac{1}{T}\int_{0}^{T}D(\rho_{S}(t),\tilde{\rho}_{S}
(t))dt\geq 1-6\sqrt{
\delta},$$ where $\rho_{S}(t)$ is the reduced state of $S$ under the
evolution of the system with initial state $\ket{\pi_{S}}\otimes\ket{\pi_{R}}$, 
$\tilde{\rho}_{S}(t)$ is the
reduced state of $S$ under the evolution of the system with initial
state $\ket{\tilde{\pi}_{S}}\otimes\ket{\pi_{R}}$, and
$D(\rho_{S}(t),\tilde{\rho}(t))=\frac{1}{2}||\rho_{S}-\tilde{\rho}_{S}||_{1}$ is the trace
distance. It is thus shown that if the dynamics of the global system is localized in state-space then it is guaranteed that large enough subsystems composed of $N_S$ sites do not equilibrate. This means that correlations do not extend over $N_S$ sites and thus the correlation length should be much smaller.

\emph{Proof}. Take a subsystem with $N_{S}$ sites satisfying
$\frac{1}{d^{N_{S}}}<\frac{\delta}{1+\delta}$. Assume, by contradiction, 
that for each
product vector $\ket{\pi_{R}}$ on the remaining system $R$, there is at most 
one product state
$\ket{\pi_{S}}$ such that $\ket{\pi_{S}}\otimes\ket{\pi_{R}}\in \Pi$ and
$L(\ket{\pi_{S}}\otimes \ket{\pi_{R}})> 1-\delta$. We would therefore have 
at least $d^{N-N_{S}}\times(d^{N_{S}}-1)$ elements of $\Pi$ 
satisfying
$L(\ket{\pi})\leq 1-\delta$. But then
\begin{align}
\braket{\braket{\overline{L}}}&=\frac{1}{d^{N}}\sum_{\pi}L(\ket{\pi})= 
\frac{1}{d^{N}}[\sum_{\pi:L(\ket{\pi})\leq 1-\delta}L(\ket{\pi})+\sum_{\pi:L(\ket{\pi})> 1-\delta}L(\ket{\pi})]\\
&\leq \frac{1}{d^{N}}[(1-\delta).(d^{N_{S}}-1).d^{N-N_{S}}+1.d^{N-N_{S}}]\\
&=(1-\delta)(1-\frac{1}{d^{N_{S}}})+\frac{1}{d^{N_{S}}}\\
&\leq (1-\delta)(1-\frac{\delta}{\delta+1})+\frac{\delta}{\delta+1}\\
&=\frac{1}{1+\delta},
\end{align}
a contradiction.

From the argument above we must have then two orthogonal product vectors 
$\ket{\pi_{s}}$ and
$\ket{\tilde{\pi}_{s}}$ on the state space of $S$ and a vector $\ket{\pi_{R}}$ 
on the space of $R$ such
that $L(\ket{\pi_{S}}\otimes \ket{\pi_{R}})>1-\delta$ and 
$L(\ket{\tilde{\pi}_{S}}\otimes
\ket{\pi_{R}})>1-\delta$. 

Let $\ket{\pi(t)}$ be the evolved state of an element $\ket{\pi}$ of $\Pi$. 
Since we can write, for every $t\geq 0$
$$\ket{\pi(t)}=\alpha(t)\ket{\pi}+\beta(t)\ket{\psi(t)}\qquad 
\text{and}\qquad \ket{\tilde{\pi}(t)}=\tilde{\alpha}(t)\ket{\tilde{\pi}}+\tilde
{\beta}(t)\ket{ \tilde{\psi}(t)},$$
where $\alpha(t)=\braket{\pi|\pi(t)}$, $\ket{\psi(t)}$ is normalized and 
orthogonal to $\ket{\pi}$ for
all $t$ (so $|\alpha(t)|^{2}+|\beta(t)|^{2}=1$), and analogous definitions for 
$\ket{\tilde{\pi}(t)}$, we
have:
\begin{align}
D(\rho_{S}(t),\tilde{\rho}_{S}(t))&=D(\text{Tr}_{R}\ket{\pi(t)}\bra{\pi(t)},
\text {Tr}_{R}\ket{\tilde{\pi}(t)}\bra{\tilde{\pi}(t)})\\
&=\frac{1}{2}||\text{Tr}_{R}[|\alpha(t)|^{2}\ket{\pi}\bra{\pi}+|\beta(t)|^{2}
\ket{\psi(t)}\bra{\psi(t)}+(\alpha(t)\beta(t)^{*}\ket{\pi}\bra{\psi(t)}+h.c.)\\
&-|\tilde{\alpha}(t)|^{2}\ket{\tilde{\pi}}\bra{\tilde{\pi}}-|\tilde{\beta}(t)|^{2}
\ket{\tilde{\psi}(t)}\bra{\tilde{\psi(t)}}-(\tilde{\alpha}(t)\tilde{\beta}(t)^{*}\ket{\tilde{\pi}}
\bra{\tilde{\psi}(t)}+h.c.)||_{1}\\
&=\frac{1}{2}||\ket{\pi_{S}}\bra{\pi_{S}}-\ket{\tilde{\pi}_{S}}\bra{\tilde{\pi}_{S}}-|\beta(t)|^{2}\ket{\pi_{
S}}\bra{\pi_{S}}+|\tilde{\beta}(t)|^{2}\ket{\tilde{\pi}_{S}}\bra{\tilde{\pi}_{S}}\\
&+\text{Tr}_{R}[|\beta(t)|^{2}\ket{\psi(t)}\bra{\psi(t)}+(\alpha(t)\beta(t)^{*}\ket{\pi}\bra{\psi(t)}+h.c.)\\
&-|\tilde{\beta}(t)|^{2}\ket{\tilde{\psi}(t)}\bra{\tilde{\psi}(t)}-(\tilde{\alpha}(t)\tilde{\beta}(t)^{*}\ket{\tilde{\pi}}
\bra{\tilde{\psi}(t)}+h.c.)]||_{1},
\end{align}
were in the last equality we have taken the partial trace of projectors $\ket{\pi}\bra{\pi}$ and
$\ket{\tilde{\pi}}\bra{\tilde{\pi}}$ and used 
that $|\alpha(t)|^{2}+|\beta(t)|^{2}=1$ and $|\tilde{\alpha}(t)|^{2}+|\tilde{\beta}(t)|^{2}=1$. Using the reverse 
triangle inequality in the last expression, we have
\begin{align}
D(\rho_{S}(t),\tilde{\rho}_{S}(t))&\geq
\frac{1}{2}||\ket{\pi_{S}}\bra{\pi_{S}}-\ket{\tilde{\pi}_{S}}\bra{\tilde{\pi}_{S}}
||_{1}-\frac{1}{2}||-|\beta(t)|^{2}\ket{\pi_{
S}}\bra{\pi_{S}}+|\tilde{\beta}(t)|^{2}\ket{\tilde{\pi}_{S}}\bra{\tilde{\pi}_{S}}\\
&+\text{Tr}_{R}[|\beta(t)|^{2}\ket{\psi(t)}\bra{\psi(t)}+(\alpha(t)\beta(t)^{*}\ket{\pi}\bra{\psi(t)}+h.c.)\\
&-|\tilde{\beta}(t)|^{2}\ket{\tilde{\psi}(t)}\bra{\tilde{\psi}(t)}-(\tilde{\alpha}(t)\tilde{\beta}(t)^{*}
\ket{\tilde{\pi}}\bra{\tilde{\psi}(t)}+h.c.)]||_{1}\\
&\geq 1-|\beta(t)|^{2}-|\tilde{\beta}(t)|^{2}-2|\alpha(t)||\beta(t)|-2|\tilde{\alpha}(t)||\tilde{\beta}(t)|\\
&\geq 1-3|\beta(t)|-3|\tilde{\beta}(t)|. \label{ineq1}
\end{align}

Now, since the Hamiltonian is non-degenerate, we have the well-known relation:
\begin{align}
\lim_{t\rightarrow\infty}\frac{1}{T}\int_{0}^{T}|\alpha(t)|^{2}&=\lim_{
t\rightarrow\infty}\frac{1}{T}\int_{0}^{T}|\braket{\pi|\pi(t)}|^{2}\\&=\lim_{
t\rightarrow\infty}
\frac { 1}{T}\int_{0}^{T}|\sum_{E}|\braket{\pi|E}|^{2}e^{-iEt}|^{2}\\
&=\lim_{t\rightarrow\infty}\frac{1}{T}\int_{0}^{T}\sum_{E,E'}|\braket{\pi|E}|^{2}
|\braket{\pi|E'}|^{2}e^
{ -i(E-E')t}\\&=\sum_{E}|\braket{\pi|E}|^{4}\\
&=L(\ket{\pi}),
\end{align}
and similar one for $L(\ket{\tilde{\pi}})$. Moreover, by the Cauchy-Schwarz inequality we have 
\begin{align}
\lim_{T\rightarrow\infty}\frac{1}{T}\int_{0}^{T}|\beta(t)|dt 
&\leq\lim_{T\rightarrow\infty}\frac{1}{T}\sqrt{\int_{0}^{T}|\beta(t)|^{2}dt}
\sqrt{ \int_{0}^{T}1dt}\\
&=\lim_{T\rightarrow\infty}\sqrt{\frac{1}{T}\int_{0}^{T}|\beta(t)|^{2}dt}\\
&=\sqrt{1-L(\ket{\pi})}\leq\sqrt{\delta},
\end{align}
and similar inequality for $\tilde{\beta}(t)$. Finally, from Inequality~\eqref{ineq1}, we get
$$
\lim_{T\rightarrow\infty}\frac{1}{T}\int_{0}^{\infty}D(\rho_{S}(t),\tilde{\rho}_{S}(t))\geq 
1-6\sqrt{\delta}. 
$$

\twocolumngrid

\end{document}